\documentclass{emulateapj}

\usepackage{graphicx}
\usepackage{epstopdf}
\usepackage{amsmath}
\usepackage{amssymb}
\usepackage{bm}

\def\gs{\mathrel{\raise0.35ex\hbox{$\scriptstyle >$}\kern-0.6em
\lower0.40ex\hbox{{$\scriptstyle \sim$}}}}
\def\ls{\mathrel{\raise0.35ex\hbox{$\scriptstyle <$}\kern-0.6em
\lower0.40ex\hbox{{$\scriptstyle \sim$}}}}

\newcommand{\mbh}{$M_{\rm BH}$}

\shorttitle{The mass of the black hole in Arp 151}
\shortauthors{Brewer et al.}

\begin{document}

\title{The Mass of the Black Hole in Arp 151 \\from Bayesian Modeling of Reverberation Mapping Data}

\author{Brendon J. Brewer\altaffilmark{1}, Tommaso Treu\altaffilmark{1}, Anna Pancoast\altaffilmark{1},\\ Aaron J.~Barth\altaffilmark{2}, Vardha N.~Bennert\altaffilmark{1}, Misty C.~Bentz\altaffilmark{3}, Alexei V.~Filippenko\altaffilmark{4},\\Jenny E.~Greene\altaffilmark{5}, Matthew A.~Malkan\altaffilmark{6}, and Jong-Hak~Woo\altaffilmark{7}}
\email{brewer@physics.ucsb.edu}

\altaffiltext{1}{Department of Physics, University of California, Santa Barbara, CA, 93106-9530, USA}
\altaffiltext{2}{Department of Physics \& Astronomy, 4129 Frederick Reines Hall, University of California, Irvine, CA 92697-4575}
\altaffiltext{3}{Department of Physics and Astronomy, Georgia State University, Atlanta, GA 30303}
\altaffiltext{4}{Department of Astronomy, University of California, Berkeley, CA 94720-3411}
\altaffiltext{5}{Department of Astronomy, University of Texas, RLM 16.228, Austin, TX 78712}
\altaffiltext{6}{Department of Physics and Astronomy, University of California, Los Angeles, CA 90095-1547}
\altaffiltext{7}{Department of Physics \& Astronomy, Seoul National University, 599 Gwanak-ro, Gwanak-gu, Seoul, 151-742, Korea}

\begin{abstract}
Supermassive black holes are believed to be ubiquitous at the centers
of galaxies. Measuring their masses is extremely challenging yet
essential for understanding their role in the formation and evolution
of cosmic structure. We present a direct measurement of the mass of a
black hole in an active galactic nucleus (Arp 151) based on the motion
of the gas responsible for the broad emission lines. By analyzing and
modeling spectroscopic and photometric time series, we find that the
gas is well described by a disk or torus with an average radius
of $3.99\pm1.25$ light days and an opening angle of 68.9$^{+
21.4}_{- 17.2}$ degrees, viewed at an inclination angle of $67.8 \pm 7.8$ degrees (that is, closer to face-on than edge-on). The black hole mass is inferred to be $10^{6.51
\pm 0.28}~ {\rm M}_{\odot}$. The method is fully general and can be
used to determine the masses of black holes at arbitrary distances,
enabling studies of their evolution over cosmic time.
\end{abstract}


\section{Introduction}
In the past decade it has become clear that supermassive black holes
are a fundamental ingredient of the universe \citep{FxF05}. Accretion
onto their deep gravitational potential is responsible not only for
some of the most powerful sources of light
\citep{1969Natur.223..690L}, but also appears to be a key ingredient
for the formation and evolution of galaxies
\citep{Graxx04,Croxx06}. Energy released by the accretion mechanism is
believed to play a role in regulating the heating and cooling of
interstellar gas and therefore the formation of stars. The ``smoking
gun'' of this connection between galaxies and black holes is the tight
correlation between the mass of black holes and the stellar velocity
dispersion $\sigma_*$ of their host galaxies, observed at small 
redshifts \citep{FxM00,Gebxx00}.  This correlation represents the
endpoint of the so-called coevolution of galaxies and black holes,
though it is still unclear how galaxies and black holes coevolve across
cosmic time. Both galaxies and black holes are believed to grow by
mergers and accretion from initial perturbations of the density field
in the early universe; however, it is not yet known whether galaxies and
black holes grow in lockstep, or one of the two forms first and
subsequently acts as a seed for the other.

The main challenge in mapping the coevolution across cosmic time is
determining black hole masses. Traditional methods rely on spatially
resolved kinematics of stars and gas within the gravitational sphere
of influence of the black hole. Therefore, they are only applicable
with current technology in the very local universe
\citep{FxF05}. Alternative methods are needed to measure black hole
masses out to distances of several billion light years --- lookback
times corresponding to a sizable fraction of the present-day age of the universe \citep{Komxx10}.

Reverberation mapping is the most promising method for measuring the
masses of black holes powering active galactic nuclei (AGNs) at
cosmologically interesting distances \citep{BxM82, Pet93, Petxx04}.
The technique is made possible by the temporal variations in the intrinsic
brightness of the central continuum source and by the subsequent response of 
line-emitting gas well within the gravitational sphere of influence of the
black hole, known as the broad-line region (BLR). By measuring the time
delay (or {\it lag}) $\tau$ between the variations of the continuum
and the variations of the broad emission lines, the physical size of
the BLR can be determined. In addition, the typical
orbital velocity of the broad-line gas can be measured from the width
of the broad lines themselves, $\sigma_l$. Combining this velocity
measurement with the radius yields an estimate of the mass of the
central black hole, \mbh $= f \sigma_l^2 c \tau/G$ \citep{Petxx04},
where $f$ is the so-called virial coefficient, $c$ is the speed of
light, and $G$ is the gravitational constant.

Despite the simplicity of the reverberation mapping idea, its
practical implementation is beset with numerous difficulties
\citep[e.g.,][]{Kro01}. First, in its standard implementation, the formula
connecting black hole mass to spectral line width and the time lag
includes a virial coefficient that depends on the unknown geometry of
the orbiting BLR gas \citep{Onkxx04,Wooxx10,Grexx10a,DDT10}. Second,
the adopted approach is usually indirect: the data are used to measure
properties of the transfer function describing the distribution of
lags, or simply the mean lag. To constrain properties of the BLR gas
distribution itself, another layer of modeling is necessary to reveal
which possible BLR geometries are consistent with the inferred
transfer function \citep{Benxx10}.

We overcome these difficulties by applying a new framework to modeling
high-quality spectrophotometric monitoring data of the Type 1 active
nucleus of Arp 151. The combination of dynamical models with
high-quality data allows us to characterize the structure of the BLR
and achieve a direct determination of the mass of a supermassive black
hole using reverberation mapping. Our measurement is independent of
any external information on the virial coefficient.

\section{Data and Modeling Framework}

\begin{figure}
\begin{center}
\includegraphics[scale=0.45]{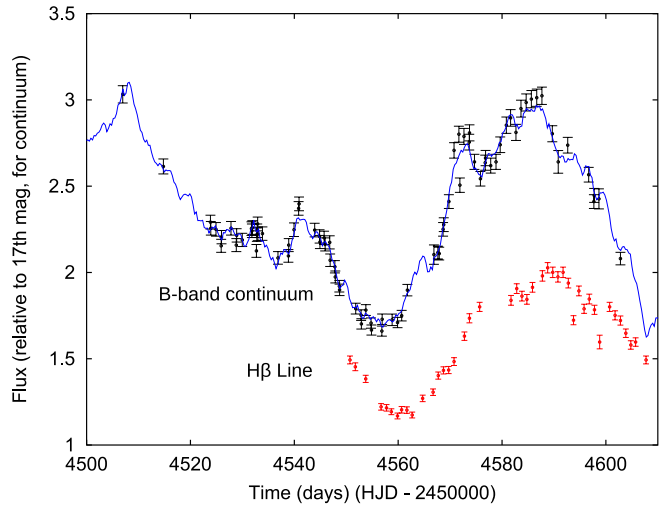}
\caption{Continuum flux time series from LAMP, and the corresponding
  H$\beta$ flux time series. The curve drawn through the continuum
  data shows a realization of an interpolation of the continuum data
  using Gaussian processes.\label{data}}
\end{center}
\end{figure}

\begin{figure}
\begin{center}
\includegraphics[scale=0.7]{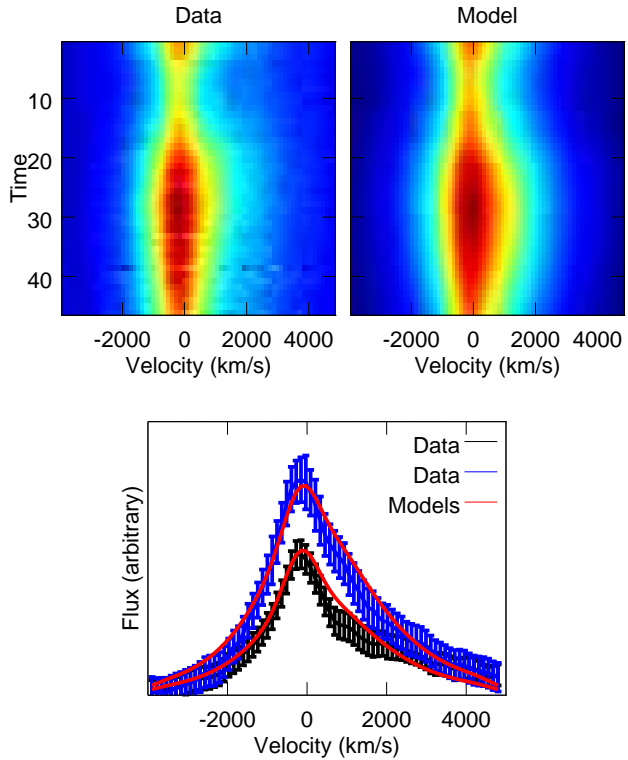}\caption{{\it Top left}:
  Measured spectrum of the broad H$\beta$ emission line as a function
  of epoch; these are the data used for our inference. {\it Top right}:
  Model-predicted spectrum as a function of time, using parameter
  values chosen at random from the posterior distribution. The major
  features of the data (time variation, line widths) are reproduced by
  our model. {\it Bottom}: Example of the spectral line shape at two
  times, along with model fits.\label{data2}}
\end{center}
\end{figure}

The data were collected as part of the Lick AGN Monitoring Project
(LAMP) campaign \citep{Benxx09,Walxx09}. They consist of 84 epochs of
photometric monitoring and 43 epochs of spectroscopic monitoring of
the region containing the H$\beta$ emission line
\citep{Benxx09,Walxx09}. The continuum and line flux temporal series
are shown in Figure~\ref{data}. The H$\beta$ spectral time series is illustrated
in the left panel of Figure~\ref{data2} (wavelength on the abscissa, epoch on the
ordinate; the epochs are approximately one day apart, but not exactly, due
to gaps and scheduling issues). 

In order to infer the BLR geometry and the black hole
mass we take a Bayesian Inference approach to the problem
\citep{sivia06}. We construct a model characterized by a finite number
of parameters describing the black hole mass, the spatial density
profile of the BLR and its kinematic structure, and the
intrinsic continuum light curve, denoting these parameters
collectively by $\Phi$. We define broad prior probability
distributions for $\Phi$ and then define the probability distribution
for the data, conditional on knowledge of all of these properties
$p(D|\Phi)$. Given specific data, the prior distribution gets updated
to the {\it posterior distribution} which describes knowledge of the
parameters after taking into account the data, using Bayes' rule
$p(\Phi|D=D^*) \propto p(\Phi)p(D|\Phi)|_{D=D^*}$. In practice, we
quantify our results by generating randomly sampled models from the
posterior distribution using a Nested Sampling method \citep{dnest}.

\begin{figure*}
\begin{center}
\includegraphics[scale=0.75]{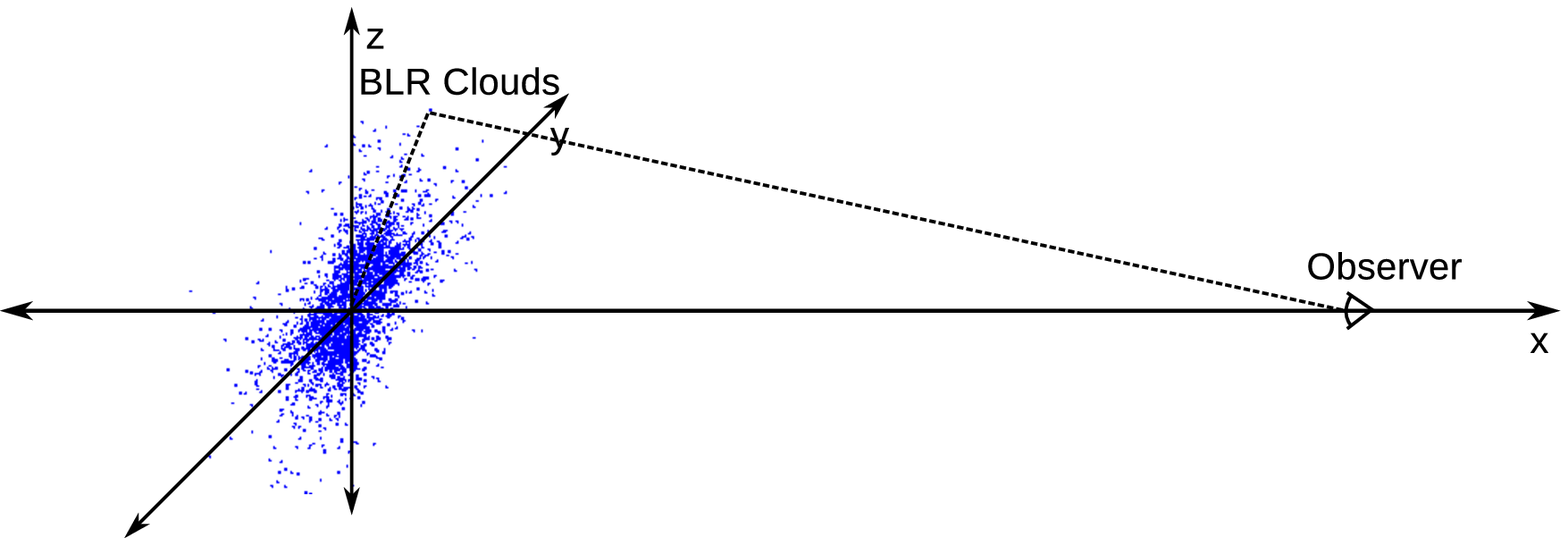}\caption{The distribution of extra path lengths
  the light must travel from the central engine to a BLR cloud and
  then to the observer is the cause of the delayed response of the
  emission-line flux, and the variations in line shape. The distribution of BLR gas in this
  diagram corresponds to a probable configuration inferred from the
  Arp 151 data.\label{diagram}}
\end{center}
\end{figure*}

The physical model consists of a large number (1,000) of BLR clouds
that are in orbit in the Keplerian potential of the central black
hole. The spatial distribution of the clouds is generated
using a flexible model that is capable of representing
generic geometries including thin disks and tori as
well as complete spheres and shells. By applying spatially varying
illumination to the clouds, we can also describe non-axisymmetric
geometries. The cloud emission is assumed to respond linearly to
continuum variations. Therefore, the observed spectrum at a given time
is the result of the continuum emission at earlier times, with time
lags corresponding to the optical path from the central continuum to
the cloud to the observer. The configuration of a model that well
represents the data is illustrated in Figure~\ref{diagram}.

Fitting of models to the data requires knowledge of the continuum flux
at all times, not just the measured times. To solve this problem, our
method uses Gaussian processes to interpolate and extrapolate the
continuum light curve taking errors into account. A typical intrinsic
light curve generated by this process is shown in Figure~\ref{data}.  Thus, our
results include uncertainty caused by the fact that we have noisy
measurements of the continuum flux at a finite number of times. Our
modeling of the continuum light curve is similar to that independently
developed by \citet{ZKP10} and used outside of reverberation mapping in studies of quasar variability \citep[e.g.,][]{2010ApJ...708..927K, KBS09, 2010ApJ...721.1014M}. Additional information on the method in general is given by 
\citet[][hereafter P11]{PBT10}.

\section{The Dynamical Model}
We do not aim to infer the position and velocity of each cloud from the
data, but rather to use the clouds to map out a spatial and dynamical
model described by a small number of hyperparameters.
This approach is equivalent to that followed by
P11, except that we are using Monte Carlo samples of clouds to
represent the distribution of BLR gas as opposed to computing the density on a
spatial grid. To generate a 3D distribution of BLR clouds, we start by generating an
axisymmetric distribution in the $x$-$y$ plane, and then apply rotations
to ``puff up'' the model into a 3D configuration. Finally, we weight the
clouds by a non-axisymmetric illumination function to model non-axisymmetric
distributions of gas.

The distance of a cloud from the black hole is prescribed according
to $r = F\mu + (1-F)\mathcal{G}$, where $F \in [0,1]$ and $\mathcal{G}$ is drawn from a gamma distribution
with mean $\mu$ and standard deviation $\beta\mu$. With
this prescription $\mu$ is the overall mean radius, $F$ is the fraction
of the mean radius that is due to the hard lower limit, and
$\beta \in [0, 1]$ describes the shape of the distribution; $\beta \approx 1$
is an exponential distribution, and $\beta \approx 0$ is a narrow normal
distribution. The polar coordinate $\phi$ of a cloud is chosen uniformly from
$[0, 2\pi]$.

For the inference, the priors on these parameters are as follows. We use a uniform prior for $\beta$, a scale-invariant $\propto 1/x$ prior for $\mu$ (between generous limits), and a uniform prior for black hole mass given $\mu$, such that the predicted line widths
are on the order of those in the data, reducing the volume of parameter space that needs to be explored.

We then assign cloud velocities in a probabilistic manner (note that
we can assign multiple velocities to each cloud in order to improve
sampling of the phase space in a computationally efficient way;
throughout this paper we adopt 100 velocities for each of the 1,000
clouds). As in P11, we assume that the only force acting on
the BLR clouds is gravity from the central black hole.

The total energy of a cloud at a distance $r$ from the black hole,
moving with angular momentum $L$, is given by
\begin{equation}
E = \frac{1}{2}\left(\dot{r}^2 + \frac{L^2}{r^2}\right) - \frac{GM}{r},
\end{equation}
which has the minimum possible value
\begin{equation}
E_{\rm min} = - \frac{GM}{r}.
\end{equation}
If we knew the position, energy, and angular momentum of a cloud, we 
could solve for the radial velocity,
\begin{equation}
\dot{r} = \pm \sqrt{2\left(E + \frac{GM}{r}\right) - \frac{L^2}{r^2}}.
\end{equation}
We choose the negative (inbound) solution with probability $q$ and the outbound with probability $1-q$, a free parameter.
For solutions to exist, the angular momentum must satisfy
\begin{equation}
L^2 \leq L_{\rm max}^2 = 2r^2\left(E + \frac{GM}{r}\right).
\end{equation}

\begin{figure*}
\begin{center}
\includegraphics[scale=0.5]{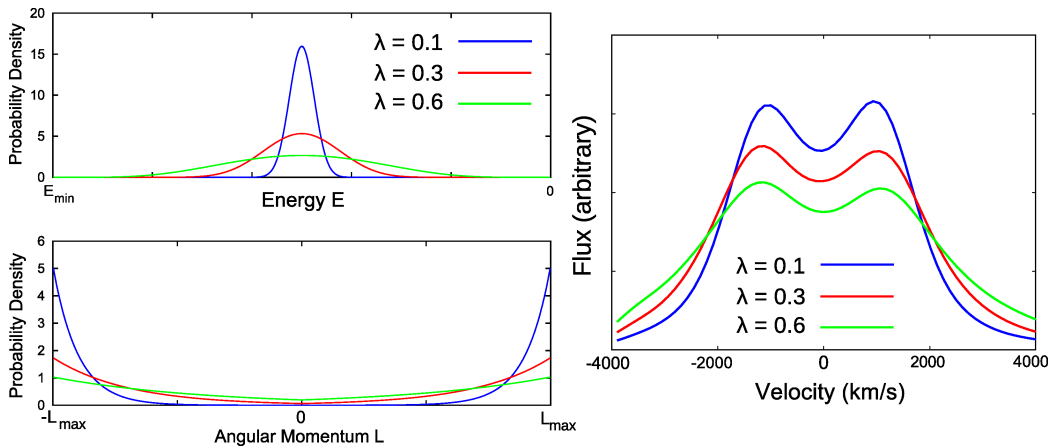}\caption{Probability distributions
  for energy and angular momentum that generate elliptical orbits for
  the clouds. The effects on the predicted line shape are illustrated
  for an edge-on disk for three different values of the
  ``noncircularity parameter'' $\lambda$.\label{EL}}
\end{center}
\end{figure*}

Circular orbits are obtained if we set the energy and angular momentum to
\begin{eqnarray}
E_{\rm circ} &=& -\frac{1}{2}\frac{GM}{r}~{\rm and} \\
L_{\rm circ} &=& \pm L_{\rm max}.
\end{eqnarray}
To get elliptical orbits, instead of assigning $E$ and $L$ the exact
circular values above, we assign them at random from the following
probability distributions:
\begin{eqnarray}
E = \left(\frac{1}{1 + \exp(-\chi)}\right)E_{\rm min},\\
\textnormal{where } \chi \sim \mathcal{N}(0, \lambda^2), \\
\textnormal{and} \\
p(L) \propto \exp\left(\frac{|L|}{\lambda}\right),~~ |L| < L_{\rm max}.
\end{eqnarray}
These probability distributions are centered around the values for
circular velocities, but the parameter $\lambda$ describes the
dispersion, or how noncircular typical orbits will be.  The circular
orbit formulae are reproduced when $\lambda \to 0$. The probability
distributions for $E$ and $L$ given three different values for
$\lambda$ are shown in Figure~\ref{EL}, along with the corresponding line
shapes. For the inference, we use a uniform prior on $\lambda$ between
0 and 1.

The main advantage of our implementation of this method with respect
to that of P11 is that we can generate relatively broad
distributions of $E$ and $L$ with a smaller number of parameters. The
drawback is that this model is technically nonstationary (in the case
of noncircular orbits) and would change if it were allowed to evolve
in a dynamically self-consistent way. Our model can be thought of as describing the
time-invariant illuminated part of the full phase-space distribution,
even though the underlying particles are actually flowing through the
region. We checked that this simplifying assumption does not bias our
inference on the black hole mass by analysis of the data with the code
developed by P11. The results are consistent with the ones
presented here.

We then rotate the models (positions and velocities of the clouds) by
an appropriate distribution of angles to generate an axisymmetric
distribution of angular momentum vectors.  The first rotation is about
the $y$ axis by a small random angle; the typical size of these angles
determines the opening angle of the disk or torus. We then rotate around the
$z$ axis by random angles to restore the axisymmetry of the model. Finally, we rotate
again about the $y$ axis, by the inclination angle to model the
inclination of the system with respect to the line of sight.
For the inference, we use uniform priors on the opening angle and
the inclination angle.

To obtain non-axisymmetric models, in order to reproduce the line
asymmetry, we weight each cloud by a simple spatially varying illumination
function. In spherical polar coordinates, this function is

\begin{equation}
W(r, \phi, \theta) = \frac{1}{2} + \kappa \cos \phi ,
\end{equation}
where $\kappa \in [-\frac{1}{2}, \frac{1}{2}]$ is a free
parameter with a uniform prior. Positive $\kappa$ illuminates the front portion of the BLR,
negative $\kappa$ illuminates the back. We also implemented, as a
secondary check, a model with a different (linear in $x$) functional form for the illumination, but this model did not reproduce the data as well, although the final black hole mass estimate was similar. In predicting the observed spectra, we included the narrow-line component as a constant which does not respond to the continuum variations.

\section{Results and Conclusions}
\begin{figure*}
\begin{center}
\includegraphics[scale=0.35]{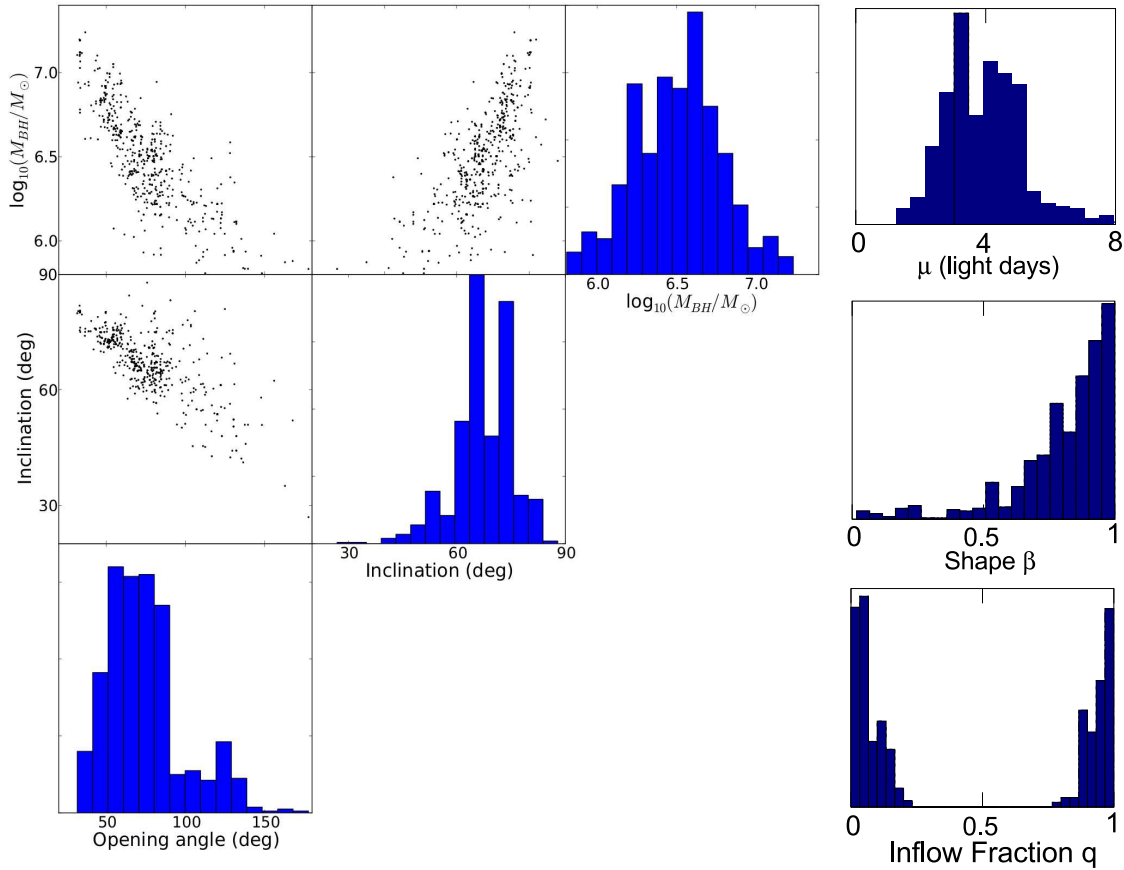}\caption{Joint and marginal posterior
  distributions for the parameters of the BLR geometry and
  kinematics. The strongest correlation is between the inclination
  angle and the opening angle of the disk. Both of these parameters
  are also strongly correlated with the black hole mass.\label{corner}}
\end{center}
\end{figure*}

Our results are presented in Figure~\ref{corner}. We find that the geometry of
the BLR is well described by a thick disk or torus (opening angle 68.9$^{+
21.4}_{- 17.2}$ degrees), viewed at an inclination $67.8 \pm 7.8$
degrees (where $0^\circ$ = edge-on and $90^\circ$ = face-on), as
depicted in Figure~\ref{diagram}. This geometry is consistent with the hypothesis
that Type 1 AGNs are viewed close to face-on, if the dusty torus is
coplanar with the BLR. The mean radius $\mu$ of the disk is
$3.99\pm1.25$ light days, and the radial profile is inferred to be
close to exponential ($\beta = 0.86^{+0.10}_{-0.19}$). We note that four light days corresponds to $<
10^{-5}$ arcsec at the distance of Arp 151 (redshift 0.021), an
angular size more than a thousand times smaller than what can be
resolved even with the {\it Hubble Space Telescope}. The orbits of the
BLR clouds are found to depart significantly from circular, and there
is no strong evidence favoring either inflow or outflow (see Figure~\ref{corner}).

We note that the posterior distribution does not rule out radial profiles
that peak close to $r=0$, which is unphysical due to the high ionizing
flux from the accretion disk \citep{2004ApJ...606..749K}. However,
conditioning on the peak of the radial profile being at the high end
of the posterior does not significantly change any inferences except
for the inflow fraction $q$: inflow becomes favored by a ratio of
70:30 if we assume the density peaks at $r > 1$ light day.
Thus, there is weak evidence for inflow, as found by \cite{Benxx10}. If inflow is present, then the front of the disk must be more
visible than the back (i.e., $\kappa > 0$), either because of obscuring
material or nonuniform illumination effects.

By marginalizing over all of the model parameters we derive the
posterior probability distribution function for the central black hole
mass. The median and 68\% credible interval are $10^{6.51
\pm 0.28}~ {\rm M}_{\odot}$.  This is lower than, but overlaps with, the value of
$10^{6.85 \pm 0.07}~ {\rm M}_{\odot}$ obtained by \citet{Benxx09}
assuming $\log_{10} f=0.74$ based on requiring active and inactive
galaxies to obey the same correlation between \mbh\ and host-galaxy
stellar velocity dispersion $\sigma_*$ \citep{Onkxx04}, and neglecting
uncertainty in $f$.  Recent measurements suggest that the intrinsic
uncertainty in $f$ from this method is at least 0.4 dex
\citep{Wooxx10,Grexx10a}. Reversing the traditional argument, our measurement implies
that $\log_{10} f=0.40\pm0.28$, a low value, for this particular system. Modeling a larger sample of systems and comparing the results of traditional methods with our direct approach
would allow us to test the assumption that active galaxies obey the same \mbh$-\sigma_*$
relation as inactive galaxies \citep{Davxx06,HxM08,Onkxx07}. 

To summarize, our measurement has three key advantages with respect to
traditional methods. First, it is direct, independent of any
assumption regarding the correlations between supermassive black holes
and their host galaxies, thus allowing us to test this assumption.
Second, it provides a more precise measurement (smaller formal uncertainties) than traditional
methods. Finally, the outputs of the inference procedure are physical properties of the BLR,
rather than a transfer function, bypassing the need for an additional modeling step.

We conclude by listing some of the limitations of this work and the
prospects for future improvements. In our dynamical
model we neglect radiation-pressure support and the dynamical influence of the accretion disk itself, as is commonly the case
in traditional reverberation mapping analysis. We also neglect the optical depth of the BLR clouds themselves \citep{1997ApJ...479..200B}. If the motion of the
BLR clouds were partially supported by radiation pressure, then we
would be underestimating the mass of the central black hole.  This is
currently a topic of debate \citep{Marxx08,NxM10}, and external
information needs to be used to break the degeneracy between black
hole mass and pressure support. Once external information is
available, it can easily be taken into account in
interpreting our results.

On a more detailed level, our model for the spatial profile of the BLR is significantly oversimplified with respect
to the real physical picture. Thus, we cannot all
features of the observed line profiles (Figure~\ref{data2}). Neglecting this
systematic uncertainty would lead us to underestimate the uncertainty
in the parameter values. In this study, we addressed this issue by inflating
the observed error bars (see Figure~\ref{data2}) until the model reproduced just the
macroscopic features of the emission lines. In other
words, we do not expect to be able to model all features of the lines,
to within the given noise level. Inflating the measurement error bars
on the data protects our results from some (but not all) systematic errors, particularly
those that would result in fluctuations in the line profiles smaller than the domain
of the data. See \citet{2011MNRAS.412.2521B} for a discussion of this issue.

For these reasons, our uncertainty in
the black hole mass is significantly larger than what the method can
in principle deliver for data of comparable quality in the absence of
modeling errors, $\sim 0.05$ dex (P11). In addition, our model
neglects collisional effects as well as anisotropic winds, which
could change somewhat the dynamics and geometry of the BLR, 
but should not affect the inference on black hole mass as long
as gravity is the dominant force.

Thus, modeling uncertainties dominate over measurement uncertainties,
driving the total uncertainty in black hole mass. Therefore, the next step toward
improving the overall precision of the measurement is to develop more
flexible and physically realistic models. Such models will also allow us to explore in more
detail the kinematics of the BLR.

\medskip
\bigskip
BJB, AP, and TT acknowledge support by the NSF through CAREER award
NSF-0642621, and by the Packard Foundation through a Packard
Fellowship. In addition, AP is funded by the NSF
Graduate Research Fellowship Program. The LAMP project was also
supported by NSF grants AST-0548198 (UC Irvine) and AST-0507450 (UC
Riverside).  AVF is grateful for the financial support of NSF grant
AST-0908886 and the TABASGO Foundation.  JHW acknowledges support by
the Basic Science Research Program through the National Research
Foundation of Korea funded by the Ministry of Education, Science and
Technology (2010-0021558).  BJB thanks Matt Auger for assistance with
plotting, and Greg Dobler and Sebastian Hoenig for useful suggestions. We thank W.~P.~McCray for useful comments that improved the clarity of the manuscript, and the anonymous
referee for suggestions that improved the modeling.

\end{document}